\documentstyle[sprocl,epsf,epsfig,wrapfig]{article}
\def\be{\begin{equation}}
\def\ee{\end{equation}}
\def\bea{\begin{eqnarray}}
\def\eea{\end{eqnarray}}

\begin{document}
\thispagestyle{empty}
\enlargethispage{0.8in}
\vspace{-2.cm}
\begin{flushright}
UCL/HEP 97-05
\end{flushright}
\vspace{0.6cm}

\title{AN EXPERIMENTER'S HIGHLIGHTS~\footnote{Invited review talk given
at Photon'97, Egmond aan Zee, the Netherlands,  10--15 May 1997}}

\author{D.J. MILLER }

\address{University College London, Gower Street, London WC1E 6BT,
England}

\maketitle\abstracts{This selection concentrates more on $\gamma \gamma$ 
results, with some reference to the related HERA photoproduction data.
Progress has been made on a wide range of topics, from $F_{2}^{\gamma}$
to the ``stickiness'' of glueball candidates, 
but many channels still need better
statistics and/or a real photon target before they can match the
comparable $ep$ studies.
}

\section{Introduction}

The HERA photoproduction data are the main experimental component in Forshaw's
``Theorist's Highlights"~\cite{Forshaw}, so this talk gives more stress to
photon-photon collisions, with some mention of related HERA results.  There are
five sections; $F_{2}^{\gamma}$ hadronic; Other Photon Structure Functions;
Inclusive Processes;  Exclusive Processes; and Dreams - possible future
developments.  Many important contributions have had to be left out for lack of
time and space.

\section{$F_{2}^{\gamma}$, hadronic}

New results on $F_{2}^{\gamma}(x,Q^{2})$ from singly tagged events have been
presented by three LEP experiments, DELPHI~\cite{Tyapkin}, ALEPH~\cite{Finch}
and OPAL~\cite{Nisius,Bechtluft}.  

 \begin{wrapfigure}[12]{r}{5.2cm}
 \vspace{-0.2cm}
 \epsfig{file=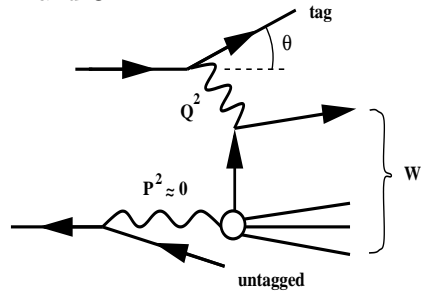,height=3.8cm,width=5.5cm}
 \vspace{-0.7cm}
 \caption{Variables in electron-photon DIS.}
 \label{fig:feynman}
 \end{wrapfigure}
Figure~\ref{fig:feynman} is the Feynman graph for a $\gamma \gamma$ scattering
event at an $e^{+}e^{-}$ collider.  For singly tagged events one of the
scattered electrons is detected, giving a good measurement of
$Q^{2}=2E_{b}E_{tag}(1-cos \theta_{tag})$ for the probing photon.  The other
lepton is required not to be seen, which keeps the value of $P^{2}$ for the
target photon close to zero.  The invariant mass of the hadronic system is
$W_{\gamma \gamma}$, which is underestimated because some of the hadronic
energy is poorly measured in the forward regions of the 
detectors~\cite{Lauber,OPALZeits}.  This means that 
the value of $x=Q^{2}/(Q^{2}+W_{\gamma
\gamma}^{2})$ is overestimated.  The experiments use unfolding
packages~\cite{Blobel,Kartvili} to correct for this.  (Things are much easier
at HERA for the measurement of the proton structure function.  There the target
proton has a unique high momentum instead of the soft distribution of virtual
target gammas radiated from the electron beams at LEP, the $ep$ event rate at
large values of W is much higher than in $e \gamma$, and $x$ is well
determined.)

OPAL~\cite{Bechtluft} has new $F_{2}^{\gamma}(x,Q^{2})$ data for two bins with
average $Q^{2}$  values of 1.86 and 3.76 GeV$^{2}$, the first measurements in
this low $Q^{2}$ region since TASSO~\cite{TASSO} and TPC/2$\gamma$~\cite{TPC}.
Electrons were tagged in the OPAL Silicon-Tungsten luminometer at angles down
to 27 mrad from the beam, with the LEP $e^{+}e^{-}$ energy close to the peak of
the $Z^{0}$.  Since LEP has now moved on past the $WW$ threshold these may be
the last measurements in this $Q^{2}$ range for a long time.  

\begin{figure}[t]
\vspace{-1.3cm}
\begin{center}
\mbox{\hspace{-1.1cm}\epsfig{file=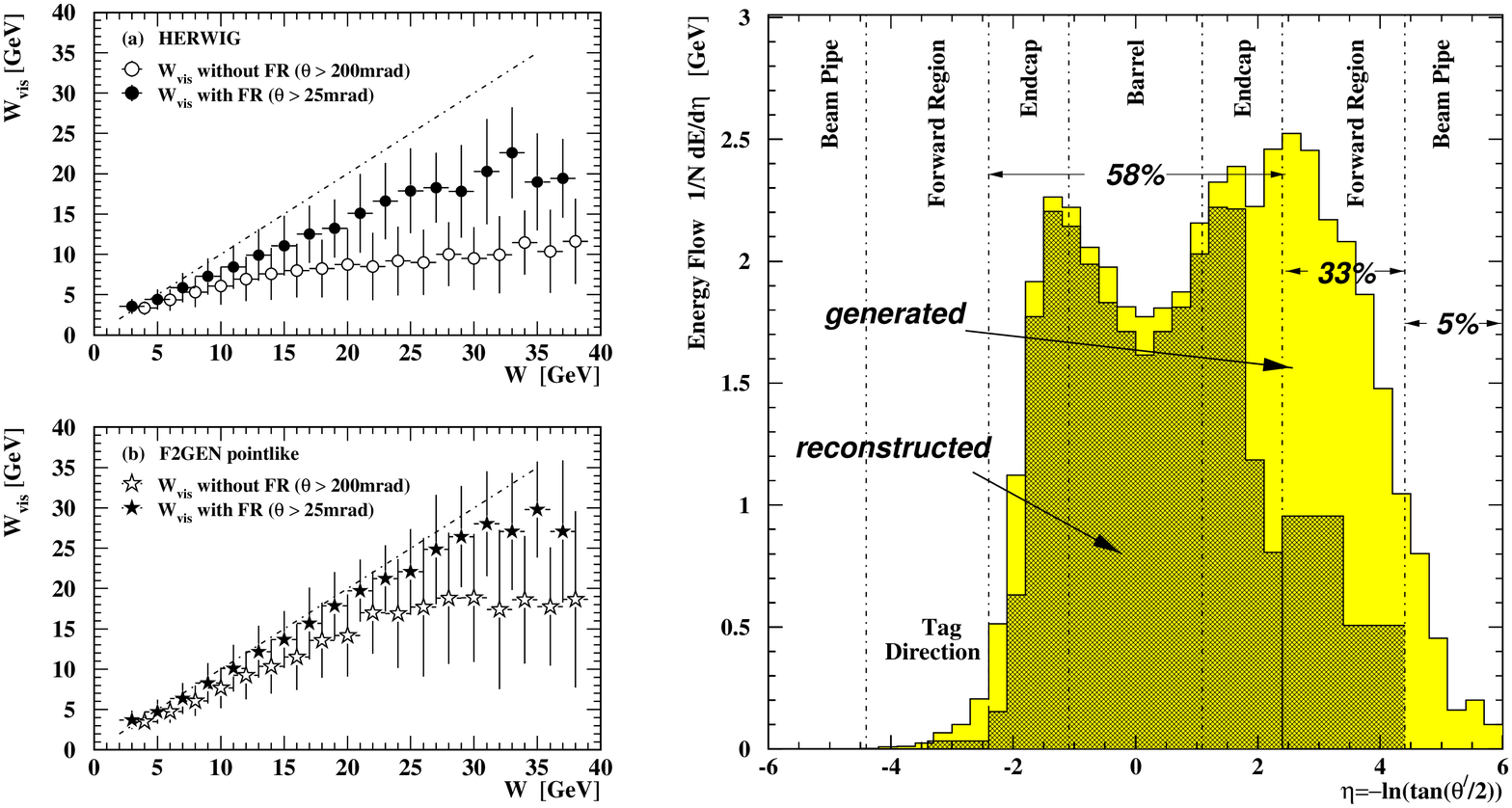,height=10cm}}

  \vspace{-2.1cm}\begin{minipage}[t]{0.42\linewidth}
  \caption{\label{fig:Wwvis} 
     $W-W_{\rm vis}$ correlation for various Monte Carlo models with
     and without the included simulation of the OPAL forward region
     (FR) between $25<\theta <200$ mrad.}
\end{minipage}\hfill
\begin{minipage}[t]{0.52\linewidth}
  \caption{\label{fig:eflow}  
     Hadronic energy flow per event as a function of pseudorapidity
     based on the HERWIG generator,
     before and after detector simulation. The tag is always at
     negative $\eta$ and is not shown.}
\end{minipage}
\end{center}
\end{figure}

The unfolded $F_{2}^{\gamma}$ distributions at low $Q^{2}$ show the following
characteristics:
\begin{itemize}
\item There is no sudden change of the shape of $F_{2}^{\gamma}(x)$ when
$Q^{2}$ drops below $5 \rm ~GeV^{2}$ (compare shape in ref~\cite{Bechtluft}
with ref~\cite{Finch} and ref~\cite{Nisius}).  This is 
in contrast with the previous measurement from $TPC/2 \gamma$. 
\item The absolute value of $F_{2}^{\gamma}$ 
(ref~\cite{Bechtluft} Fig.~2) is higher than either
the GRV~\cite{Vogt,GRV} or the SaS-1D~\cite{SaS-1D} predictions.  The GRV-HO
curve comes closest.
\item A rise at $x<0.01$, as seen in the proton structure at
HERA~\cite{ZEUSfirst,H1first}, is allowed but not established, 
largely because -- 
\item The systematic errors after unfolding are much larger than the
statistical errors (true for all LEP $F_{2}^{\gamma}$ measurements, see
discussion in next few paragraphs).
\end{itemize}

The values of  $F_{2}^{\gamma}$ in the medium to large $Q^{2}$ range
($5<Q^{2}<120 \rm ~GeV^{2}$) from the three LEP 
experiments~\cite{Tyapkin,Finch,Nisius} are in good agreement 
(see Figure 3 in~\cite{Nisius}).  All of them
are consistent with the expected $\ln Q^{2}$ rise from QCD~\cite{Vogt}.  The
DELPHI error bars are less than those from ALEPH and OPAL, for comparable
statistics, because DELPHI has a different approach to calculating the
systematic errors from unfolding; what Lauber~\cite{Lauber} calls ``the
problem".

 \begin{wrapfigure}[17]{r}{5.9cm}
 \vspace{-0.1cm}
 \epsfig{file=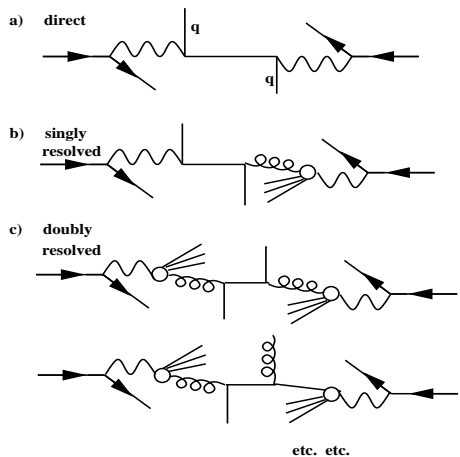,height=6.cm,width=6.cm}
 \vspace{-0.7cm}
 \caption{Feynman graphs of direct and resolved processes.}
 \label{fig:directetc}
 \end{wrapfigure}
The problem was posed -- in exaggerated form, we now know -- by Forshaw
(reporting an exercise of Seymour, corroborated by L\"onnblad) at 
Photon'95~\cite{Forshaw95}.  Events from the HERWIG~\cite{HERWIG} Monte Carlo
program were passed through a simple detector simulation which modelled the
way the experimental analyses had previously been done by suppressing all
hadron reconstruction in the endcap regions ($\theta<200 mr$).  In this HERWIG
exercise, for generated values of $W_{\gamma \gamma}>15 \rm ~GeV$ almost all
correlation was lost between the visible reconstructed value 
$W_{\rm vis}$ and the generated value -- see the open circles in
Figure~\ref{fig:Wwvis}~\cite{LauberWw}.  Studies with
PYTHIA~\cite{PYTHIA} and ARIADNE~\cite{ARIADNE} showed a similar effect.  If
this were representative of what is really happening in experiments it must
mean that, for large $W_{\gamma \gamma}$ and hence for small $x$, unfolding
results would be unreliable -- as experimenters already feared~\cite{DJMLund}.
An immediate partial remedy was clear to the experimenters; use the sampled
hadron energy from the forward electromagnetic calorimeters.  
Figure~\ref{fig:eflow} shows
that approximately one third of this energy is actually measured by OPAL.
ALEPH and DELPHI are similar.  The result is shown as the solid circles in
Figure~\ref{fig:Wwvis}(a).  Some correlation is already restored.

But study of the data has led all three LEP experiments to doubt the
completeness of the modelling in HERWIG and PYTHIA.  The measured hadronic
energy flows in OPAL and ALEPH, as reported here~\cite{Finch,Nisius}, show
less energy in the partially sampled forward region than predicted by these two
Monte Carlo models, and more energy goes into parts of the well-measured
central region.  In OPAL the shape of the observed energy flow is closer to
that from the simple F2GEN~\cite{F2GEN} model where the outgoing hadronic
system is generated as the pointlike production of a quark-antiquark pair,
though this must be an incomplete model of the QCD process.
Figure~\ref{fig:Wwvis}(b) shows how much better the correlation is
between $W_{\rm vis}$ and the true value for events generated with this pointlike
F2GEN model, both with and without the sampled hadronic energy from the forward
region.  The distribution of hadronic transverse energy $E_{\rm t,out}$,
perpendicular to the beam-tag plane, is also very different between data and
HERWIG or PYTHIA, especially at low $x$~\cite{Lauber}.  And Rooke~\cite{Rooke}
has shown that the number of events with 2 high transverse energy jets is much
lower in HERWIG and PYTHIA than in the data.  In both of these cases the
pointlike F2GEN sample lies on the other side of the data points from HERWIG
and PYTHIA.  Butterworth (private communication) has speculated that HERWIG and
PYTHIA may be underestimating the contribution from one or more hard-parton
processes; photon-gluon fusion, for instance (Figure~\ref{fig:directetc}(b)).

 \begin{wrapfigure}[17]{r}{6.9cm}
 \epsfig{file=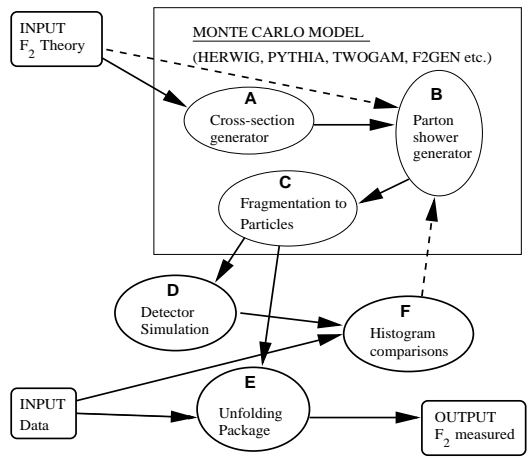,height=6.cm,width=6.9cm}
 \vspace{-0.5cm}
 \caption{$F_2^\gamma$ measurement flow chart.}
 \label{fig:flow}
 \end{wrapfigure}
The way the game is now being played is shown in Figure~\ref{fig:flow} as a
flowchart.  Lauber~\cite{Lauber} described an exercise with Seymour and
L\"onnblad which is represented by the nearly vertical dotted arrow from
item F to item B on the
flowchart, using the energy flow and $E_{\rm t,out}$ histograms from experiment,
item F, to tune the parameters of the parton shower generators,
item B, in HERWIG and PYTHIA.  Tyapkin~\cite{Tyapkin} reported a similar
exercise with the DELPHI generator TWOGAM which has an explicit singly resolved
photon component~\cite{Zimin}, including photon-gluon fusion.  The nearly
horizontal dotted arrow, from INPUT to item B on the flowchart, 
represents a feature of both HERWIG
and PYTHIA which use the input set of theoretical parton density functions in
their parton shower generators as well as in the cross section generator.  
The large systematic errors on the OPAL and ALEPH unfoldings come from
assuming a set of Monte Carlo models which cover the whole range of variations
in the histogrammed quantities.  The DELPHI errors are smaller because they
only use the tuned TWOGAM Monte Carlo for unfolding.  

There is a serious dilemma here.  If we tune the generators perfectly, to match
all of the observed histograms, then it will not matter what input
parametrisation of $F_{2}^{\gamma}(x,Q^{2})$ we have used; the unfolding
package, item E, will automatically give us back the input
$F_{2}^{\gamma}(x,Q^{2})$ as our measured output.  What is needed is a set of
Monte Carlo models whose parameters  are all tied down, either by QCD theory or
by fits to other data -- hadronic scattering, HERA photoproduction, etc.  We
must use them to unfold $F_{2}^{\gamma}(x,Q^{2})$ from the visible $x$
distribution, but we must also check that they give good energy flows, jet
numbers and $E_{\rm t,out}$ distributions.  If they do not there must be something
missing from them which will have to be added in a well motivated way, or we
have to look for better models.  It is intriguing that the PHOJET Monte Carlo
model~\cite{Engel} fits some features of the untagged $\gamma \gamma$ data as
well as PYTHIA does~\cite{Buergin}.  A version of PHOJET with off-shell photons
is eagerly awaited, as are re-engineered versions of HERWIG and PYTHIA.  

The last three or four years of LEP running will double or triple the
statistics available for photon structure function analysis.  If the Monte
Carlo tools can be refined to match there is every prospect of clear answers to
two questions; can we measure $\Lambda _{QCD}$ from the high $Q^{2}$ evolution,
and is there a rise of $F_{2}^{\gamma}(x,Q^{2})$ at low $x$?  Of course, we
would also like to measure the gluon density in the photon -- but that is only
accessible directly through inclusive processes, see below.

\section{Other Photon Structure Functions}

There is no reason to expect surprises from measurements of the QED structure
functions of the photon.  A large part of our motive for studying them is to
use them as a testbed for the techniques used to extract the hadronic structure
functions.  The longitudinal hadronic structure function $F_{L}^{\gamma}$ is
particularly interesting because it should have different QCD
scaling~\cite{Witten} behaviour from $F_{2}^{\gamma}$.  But it had been shown
before LEP started up~\cite{Millergreenbook} that $F_{L}^{\gamma}$ would be
hard to measure there because of poor statistics for events with the highest
sensitivity to $F_{L}^{\gamma}$, events with low tagged electron energies.  The
difficulties are now known to be even greater due to background from fake-tags
by off-momentum electrons in the beam halo (e.g.~\cite{Zimin}).  More recently
Field and others~\cite{Field95,LEP2wkshp} have pointed out that there are
other structure functions which are akin to $F_{L}^{\gamma}$, but which can be
measured from the main sample of tagged data.

ALEPH~\cite{Brew} and OPAL~\cite{Doucet} both reported results from singly
tagged $\gamma \gamma \rightarrow \mu^{+} \mu^{-}$ samples.  The new structure
functions govern the distribution of the azimuthal angle $\chi$ between plane
of the outgoing muons and the plane of the beam and the tagged electron in the
$\gamma \gamma$ C. of M.  Both saw significant values for $F_{B}^{\gamma,
QED}$, in agreement with the QED prediction.  ALEPH also presented the first
measurement of $F_{A}^{\gamma, QED}$.  $F_{B}^{\gamma, QED}$ multiplies the
$cos 2\chi$ term in the angular distribution and $F_{A}^{\gamma, QED}$
multiplies the $cos \chi$ term.  Since the two experiments used different sign
conventions for the definition of $\chi$ it may well be that OPAL 
``folded away''
their sensitivity to $F_{A}^{\gamma, QED}$.  Successful measurement of
$F_{B}^{\gamma, QED}$ is particularly encouraging because its hadronic form has
the same parton content as $F_{L}^{\gamma}$, in the limit of massless quarks,
though it comes from a different set of helicity amplitudes.  

The task now is to try and use the outgoing jets in hadronic events in the same
way as the outgoing muons to define a $\chi$ angle.  There will be problems.
Whereas the tagged $\mu^+\mu^-$ events have a constrained fit which gives a
precisely defined final state $\gamma \gamma$ energy, the hadronic events are
very poorly defined because of the incomplete sampling of hadron energies in
the forward regions.  And only a sub-set of hadronic events has a clear two-jet
axis.  Telnov suggested that the statistics may be increased by including
untagged events in which the electron recoil plane is implicitly defined by the
overall transverse momentum of the hadronic system, but it is not clear that
this will work.  New ideas are still needed.  If we are lucky Photon '99 may
see the first analyses for the hadronic $F_{B}^{\gamma}$ and its evolution with
$Q^{2}$.

\section{Inclusive processes}

H1 continues to tease the $\gamma \gamma$ community by trying to
extract photon structure from jets in photoproduction.  
The latest study~\cite{Rick} uses an
appropriate set of cuts to get a differential cross section which they
say should be equal to the pointlike anomalous contribution to the photon
interaction $\alpha^{-1} x_{\gamma}(q+\frac{9}{4} g)$.  When this is plotted
against the $p_{T}^{2}$ of the jets it has a logarithmic rise, as would be
expected from the scale-breaking nature of the photon-quark coupling. In fact,
the rise seems to be significantly steeper than either the GRV prediction or
the observed logarithmic rise in $F_{2}^{\gamma}(Q^{2})$~\cite{Nisius}.  This
may be a hint that the gluon contribution is doing something unexpected or --
more likely I fear at this stage -- that the H1 analysis could contain
systematic effects which have not yet been understood.  It is noticeable that
H1 did not present an update at Photon '97 of the Photon '95
analysis~\cite{Erdmann} that claimed to measure the gluonic structure of the
photon, presumably because of systematic difficulties in separating the primary
signal from underlying multiple parton interactions, as described at Photon '95
by ZEUS~\cite{Butterworth}.

Progress has been made in inclusive $\gamma \gamma$ analysis, thanks to two
important factors:  a)  LEP has moved away from the $Z^{0}$ peak;  b)  the HERA
experiments have developed analysis techniques which can be applied to $\gamma
\gamma$ as well as to $\gamma p$.  Even though the integrated LEP luminosity
above the $Z^{0}$ is still only 10s of pb$^{-1}$ compared with over $100
\rm pb^{-1}$ on peak, 
the rate for collecting untagged $\gamma \gamma \rightarrow
hadrons$ is much greater than for tagged events -- and the $Z^{0}$ background
can be kept well below 10\% of the sample with reasonable cuts.

DELPHI~\cite{Zimin} presented a preliminary empirical survey of how the
properties of events evolve with $\sqrt{s_{e^+e^-}}$.  The observed cross
section, after selection cuts, rises at about 10 pb/GeV from
$\sqrt{s_{e^+e^-}}\simeq 132 \rm ~GeV$ to 
$\sqrt{s_{e^{+}e^{-}}}\simeq 172 \rm ~GeV$,
and it extrapolates back plausibly to just below the points at
$\sqrt{s_{e^+e^-}}\simeq 91 \rm ~GeV$, under the background from the $Z^{0}$.
The same TWOGAM Monte Carlo model that they use for unfolding
$F_{2}^{\gamma}$~\cite{Tyapkin} gives predicted distributions of final state
quantities, including $W_{\gamma \gamma}$, energy flow as a function of
psuedorapidity, transverse momentum of jets and number of jets.  Most of them
agree well; this home-made model seems to have a good combination of hard and
soft components.  But they draw attention to one disagreement between data and
Monte Carlo at $\sqrt{s_{e^{+}e^{-}}}\simeq 172 \rm ~GeV$, where the energy flow in
the forward region drops below the prediction in a way which is very
reminiscent of the effect seen in the OPAL tagged
data~\cite{Nisius,Bechtluft,Lauber}.

OPAL's inclusive analysis~\cite{Buergin} goes further than DELPHI, 
and may be a prototype that other
experiments could follow ($\gamma \gamma \rightarrow hadrons$ has long had as
many different analysis techniques as experiments~\cite{MillerCornell}, which
meant that no experiment could check another's results).  OPAL uses a
development of the $x_{\gamma}$ variable from HERA as an estimator of
the fraction of the target photon's momentum carried by the hard parton which
produces identified jets with high $E_{T}$.
\[x^{\pm}_{\gamma}=\frac{\sum_{jets} {E_{j} \pm p_{z,j}}}{\sum_{hadrons}
{E_{i} \pm p_{z,i}}}, \] 
where $p_{z,i}$ is the momentum of a hadron projected along the LEP beam
direction.  The $\pm$ ambiguity arises because the initial state is
intrinsically symmetric, unlike the situation at HERA, 
and either photon might be the target.
Three main categories of events with high $E_{T}$ jets are expected: direct,
singly resolved and doubly resolved (Figure~\ref{fig:directetc})~\cite{ChrLls}.
Using the PYTHIA Monte Carlo, OPAL shows that the direct sample should be very
cleanly separated from the resolved samples by requiring both $x^{+}_{\gamma}$
and $x^{-}_{\gamma}$ to be greater than 0.8.  They confirm this separation in
the experimental data for two jet events with $E_{T}>3 \rm ~GeV$ by computing an
effective parton scattering angle $\theta^*$ in the dijet C. of M. and showing
that the direct ($x^{\pm}_{\gamma}>0.8$) sample has the  expected rather flat
distribution, while the resolved samples ($x^{+}_{\gamma}$ or $x^{-}_{\gamma}$
less than 0.8) are much more forward-backward peaked, as predicted on a parton
level by lowest order QCD (and as seen in a very similar analysis of
photoproduction by ZEUS, quoted in Aurenche's introduction to the inclusive
session~\cite{Aurenche}). 

Given the evidence, at least in the two jet sample, for approximate jet-parton
duality, OPAL has compared the $E_{T}$ distribution of jets with the parton
level NLO matrix element predictions of Kleinwort and Kramer~\cite{KandK}.  The
effects of measurement errors are removed by unfolding.  The match between
theory and experiment is good for $E_{T}>5 \rm ~GeV$ and is consistent with the
predicted domination by the direct matrix element for 
$E_{T}>8\rm ~GeV$.  Aurenche
also showed how well these NLO curves matched $\gamma \gamma$ data from AMY and
TOPAZ, as well as photoproduction from H1 and ZEUS.

Comparison of the OPAL inclusive two-jet cross sections with Monte Carlo
predictions is tantalising.  For direct events ($x^{\pm}_{\gamma}>0.8$) the
PYTHIA and PHOJETS predictions agree with one another and with the data,
regardless of the set of PDFs used.  But for $x^{+}_{\gamma}$ or
$x^{-}_{\gamma}$ less than 0.8, i.e. for the resolved samples, there are some
disagreements between the two programs with the same PDFs, and large
disagreements between different PDFs in the same program.  The LAC1~\cite{LAC}
PDFs, for instance, give much too high a cross section with both programs,
surely because of too much gluon.  Better statistics and further analysis may
lead to an independent measurement of the gluon content of the photon.

The total cross section $\sigma_{\gamma \gamma}$ has been one of the worst
measured quantities in particle physics~\cite{Pancheri} (but see ``Dreams"
below).  It remains so for $W_{\gamma \gamma}<5 \rm ~GeV$, but L3 has presented
first measurements from LEP~\cite{VanRossum} with 
$5<W_{\gamma \gamma}<70 \rm ~GeV$
which are much more coherent than anything at lower energies.  They show a
significant rise over this range, consistent with the logarithmic rise seen in
hadron-hadron and $\gamma p$ cross sections.  The problem with this measurement
is an intensified version of the problem discussed above for $F_{2}^{\gamma}$,
how to correct for the lost hadronic energy in the forward region.  In the
tagged events used for the structure function some transverse momentum is
required in the hadronic system to balance the tagged electron.  But the bulk
of the events in the total cross section have no tag, and at high $W_{\gamma
\gamma}$ there must be a large fraction of diffractive events in which the
hadrons hardly have enough transverse momentum to enter the forward luminosity
detectors.  Most of these events give no trigger and the only way of allowing
for them is to use a Monte Carlo program to correct for their loss.  Rather
surprisingly the PHOJETS and PYTHIA Monte Carlo models give very similar
distributions for the $W_{\rm visible}$ distribution, including the barrel region
and the forward detectors, so the total cross section values do not change much
when unfolded with either PYTHIA or PHOJETS.  But a plot was 
shown~\cite{VanRossum}  of cluster
energies in the forward luminosity detectors alone in which there was a marked
divergence at high energies between, on the one hand, the data and the PHOJETS
prediction, which both levelled off and agreed with one another, and on the
other hand, the PYTHIA prediction which fell away much more sharply.  This is
all we know about the region where many events must be totally unseen, so it is
hard to be completely confident in the measurement until one or more of DELPHI,
ALEPH and OPAL have done a similar analysis, hopefully with a larger selection
of Monte Carlo models.

Charm production in $\gamma \gamma$ remains intractable.  The new L3 result for
the inclusive charm cross section~\cite{Andreev} agrees with the QCD
model~\cite{DKZZ}, but it is only based on 43 events at LEP1 in 
$80\rm pb^{-1}$ and
29 events at LEP2 in $20\rm pb^{-1}$, both tagged with muons 
from charm decay.  It
is frustrating to know that there are thousands of unresolved charm events
there, boosted forward by the $\gamma \gamma$ kinematics so that they cannot be
identified in the microvertex detectors.  A few more tagging channels can be
added, however, and the eventual LEP2 luminosity should give a factor of
$\simeq \times 20$, so a worthwhile test of the theory should come by Photon
'01. 

A potentially important $\gamma^* \gamma^*$ study has been suggested by Hautmann
and others~\cite{Hautmann,deRoeck} who make predictions from the high energy
limit of QCD (using the BFKL pomeron) which give a significant doubly-tagged
rate for $e^{+}e^{-} \rightarrow e^{+}e^{-}hadrons$ (approximately 1 event per
pb$^{-1}$ at LEP2 with $Q^{2}\simeq 10\rm ~GeV^{2}$).  There was some surprise that
the effect has not yet been noticed in LEP1 data, if it is there.  A few dozen
doubly tagged events have been seen.  They are routinely rejected from the
singly tagged samples of thousands of events which are used for structure
function studies.  There may just be enough of them, after inefficiencies have
been allowed for, to accommodate the new prediction.  As ever, a Monte Carlo
study of the hadronic acceptance will be needed to find out if a significant
part of the signal is being lost.  This will surely be settled by Photon '99.
Come to Freiburg to see if BFKL survives!

\section{Exclusive processes}

There is no shortage of data, but there is a serious shortage of people to work
on it.  Cleo II now has over 3fb$^{-1}$ of integrated luminosity, and we can
expect even more from the specialised beauty factory experiments, Belle in
Japan and BaBar at Stanford.  For higher mass $\gamma \gamma$ systems LEP is
accumulating worthwhile samples.  And there is no shortage of problems to be
solved, both from QCD~\cite{Brodsky} and in resonance physics where predictions
proliferate for glueballs, hybrids, molecules, 4-quark states and recurrences.
I concentrate on two beautiful results from Cleo II, supplemented by L3, and
mention a first survey from H1.

Cleo II has sufficient integrated luminosity to do a precision study on tagged
samples of $\gamma^* \gamma \rightarrow \pi^{0}, \eta$ and $\eta
'$~\cite{Savinov}.  They have recalibrated the inner edge of their tagging
detector so that they can use incompletely contained electron showers to go
down to a lower limit of $Q^{2}=1.5 \rm ~GeV^{2}$, 
joining on well for the $\pi^{0}$
with lower $Q^{2}$ data from CELLO. There is a clear difference between the
$Q^{2}$ behaviour of $\eta'$ and the behaviour of $\pi^{0}$ and $\eta$.  Both
$\pi^{0}$ and $\eta$ form factors appear to obey the perturbative QCD
prediction of Brodsky and Lepage~\cite{Brodsky}:  
\[ \lim_{Q^{2} \rightarrow \infty}|F_{\gamma^* \gamma m}(Q^{2})|=2f_{m}, \]
where $m$ is the particular pseudoscalar meson,
and they have consistent values ($\Lambda_{\pi^{0}}\simeq 776 \pm 20\rm ~MeV,
\Lambda_{\eta}\simeq 774 \pm 30$ MeV) for the $\pi^{0}$ and $\eta$ 
mass parameters in the monopole formula:
\[F(Q^{2})=F(0) \frac{1}{1+Q^{2}/ \Lambda_{m}^{2}}.\]
But the $\eta '$ form factor rises to approximately twice the pQCD prediction
at $Q^{2}\simeq 15 \rm ~GeV^{2}$, and it has a higher monopole mass ($\Lambda_{\eta
'}\simeq 859 \pm 25$ MeV; L3 is consistent~\cite{Braccini} but with bigger
errors).  Brodsky and Ruskov -- in their talks~\cite{Brodsky,Ruskov} and over
breakfast this morning -- agree that these results mean that the $\pi^{0}$ and
$\eta$ are behaving as if their wavefunctions are already close to asymptotic
whereas the $\eta '$ is a much more complicated mixed object.

Cleo II's other beautiful result was totally negative~\cite{Paar} but very
clear.  This was a search for $\gamma \gamma$ production of the glueball
candidate $f_{J}(2220)$ and its decay to $K_{s}K_{s}$.  Cleo II sees many other
resonances in this analysis, so there is no question about their sensitivity,
but they do not see even a hint of the $f_{J}(2220)$.  They therefore put the
highest ever lower limit ($>82$ at 95\% confidence) on the
``stickiness"~\cite{Chanowitz} of a meson, the normalised ratio of its $\gamma
\gamma$ width to its radiative branching ratio from $J/\psi $.  Both BES and Mk
II have clear signals for $J/\psi$ decays to 
the $f_{J}(2220)$.  This object must now be one of
the strongest of all glueball candidates.  Two other experiments,
L3~\cite{Braccini} and ARGUS~\cite{Medin} reported $\gamma \gamma$ resonance
studies.  The L3 results are promising and should soon have a physics impact.
They demonstrate a good acceptance and resolution for many states with masses
from 1200 to 1750 GeV/$c^{2}$ and the statistics will triple or quadruple
before Photon '01.  

There was an encouraging first look at exclusive resonance production at HERA
from H1~\cite{Tapprogge}, making particular use of the new SPACAL calorimeter
to measure multi photon final states boosted in the backward direction.  Clear
$\pi^{0}$, $\omega$ and $\eta$ signals were seen, but no $\eta '$.  There was
also a suggestion of an $a_{0}(980)$ peak.  As well as conventional $\gamma
\gamma$ or $\gamma $-pomeron processes, some of these channels should be
sensitive to more exotic exchanges, such as the ``odderon".  With rising HERA
luminosity this could become very interesting.

\section{Dreams; possible future developments}

A recurrent good dream seems closer to the real world after Romanov's
talk~\cite{Romanov}.  This is the hope for precise measurement of the total
cross section $\sigma_{\gamma \gamma}$ in the resonance region
by using double tagging at around zero
scattering angle in an $e^{+}e^{-}$ collider.  The KEDR detector at the VEPP-4M
collider in Novosibirsk has focusing spectrometers built into it which measure
the outgoing electron and positron to very high precision (we saw results from
a setting-up experiment on photon splitting using one of the two
spectrometers~\cite{Maslennikov}).  The collider will run with $\sqrt{s} \simeq
1$ GeV soon, but should then go up to around 12 GeV.  The resolution on the
mass of the system recoiling against the two tags will be better than 20
MeV/$c^{2}$ over a range of masses from $\simeq 0.5$ to 3.5 GeV/$c^{2}$, with a
tagging efficiency of better than 15\%.  The main KEDR detector will have good
tracking and calorimetry to measure the properties of the hadronic final state,
so this experiment could make a substantial contribution to resonance studies.
A daydream which some of us indulge in is to imagine the same kind of zero
angle tagging system installed in one of the spare LEP straight sections,
together with good luminosity monitors and forward tracking, with a simple
barrel detector to trigger on hadronic systems.  A well designed specialised
experiment could push the $\sigma_{\gamma \gamma}$ measurement up to
$\sqrt{s}\simeq 70$ GeV or more, could solve the big problem of measuring
$W_{\gamma \gamma}$ in the study of $F_{2}^{\gamma}$, could see the BFKL
effects predicted by Hautmann et al. and would be much more sensitive 
than the present LEP experiments to such
diffractive processes as $\gamma \gamma \rightarrow \rho \rho, J/\psi \rho$
etc.  But I hear there is to be a new user for the LEP tunnel after 2001.

In this morning's talks on the high energy photon linear collider Telnov
reported~\cite{Telnov} on the steady progress being made in solving the
fundamental problems of realising the full potential luminosity of such a
machine and Jikia~\cite{Jikia}, Ginzburg~\cite{Ginzburg} and
Takahashi~\cite{Takahashi} updated some of the feasibility studies on physics,
including measuring the couplings of Higgs bosons to $\gamma \gamma$.  Because
this coupling could be sensitive to the existence of very heavy fermions and
bosons -- well beyond anything reachable at 
planned machines -- it remains one of
the most important of all the numbers to be determined once a Higgs boson is
found.  Nothing has been said here to undermine the conclusion presented at the
LCWS in Morioka~\cite{DJMMorioka} that, if a Higgs boson is found with a mass
of less than 350 GeV, then a high energy $\gamma \gamma$ collider must be built
to study it.  Such a machine in $e^{-} \gamma$ mode will also give the
definitive measurement of the high $Q^{2}$ evolution of $F_{2}^{\gamma}$,
avoiding the big problem of measuring $W_{\gamma \gamma}$ by using a narrow
band beam of real photons as the target~\cite{VogtMiller}.  Brodsky says that
he believes the study of  $e^{-} \gamma \rightarrow W\nu $ will give the best
possible measurement of the $\gamma WW$ couplings.  Telnov reminded us that if
a high energy linear $e^{+}e^{-}$ collider is built there must be provision for
a second interaction region with a finite beam crossing angle to be built at a
later date for real $\gamma \gamma$ and $\gamma e^{-}$ physics.

The idea of a lower energy photon linear collider was mentioned in passing.  It
could be a superb tool for studying resonances in the 1 to 4 GeV/$c^{2}$ mass
range~\cite{Borden,MillerBerk}.  If it were done as part of an upgrade of the
SLC at Stanford it might even reach the $e^{-} \gamma \rightarrow W\nu $
threshold.

\section{Summary and Conclusions}

In measuring $F^{2}_{\gamma}$ the LEP experiments agree with one another that
the shape and evolution are consistent with QCD.  But the problem of modelling
the parton shower must be solved before the two important questions can be
settled: is the hadronic part of the photon so like the proton that at low x it
has the same kind of rising structure function; and can a precise measurement
of the QCD scale be made from the evolution at high $Q^{2}$?  The influence of
HERA photoprodcution on untagged $\gamma \gamma$ studies is very important.  It
will be intriguing to see whether LEP or HERA gets the best eventual
measurement of the gluon density in the photon; each has its own systematics
and intrinsic background problems.  Resonance studies continue to be frustrated
by lack of effort; the work is intricate and time consuming, and it can be
unrewarding if the results are not clear cut.  Here Cleo II used its large
statistics to report two convincingly clear results.  L3 should be able to
follow suit with its excellent neutral particle reconstruction.

The connections between photoproduction and $\gamma \gamma$ physics grow
closer.  Many of the ``dreams" of $e \gamma$ and $\gamma \gamma$ physicists,
from the previous section, involve achieving comparable statistics and
precision to what HERA can already do in $ep$ or $\gamma p$.  This may only
be possible at a linear collider.

\section*{Acknowledgements}
The organisers of the conference are to be congratulated on the
scientific organisation, on their
choice of venue and on the care they have taken of us.  Comletion of
the written version of this review has depended heavily upon the kind
advice and help of Dr. Jan Lauber.

\vfill\eject


\section*{References}
\begin{thebibliography}{99}

%-----------------------------------------------------------------------
\bibitem{Forshaw}
J.~R.~Forshaw.  These Proceedings.
%-----------------------------------------------------------------------
\bibitem{Tyapkin}
I.~Tyapkin for the DELPHI collaboration. These Proceedings.
%-----------------------------------------------------------------------
\bibitem{Finch}
A.~Finch for the ALEPH collaboration.  These Proceedings.
%-----------------------------------------------------------------------
\bibitem{Nisius}
R.~Nisius for the OPAL collaboration.  These Proceedings.
%-----------------------------------------------------------------------
\bibitem{Bechtluft}
J.~Bechtluft for the OPAL collaboration.  These Proceedings.
OPAL Collaboration, K.~Ackerstaff {\em et al}, Z.Phys. {\bf C74} (1997)33
%-----------------------------------------------------------------------
\bibitem{Lauber}
J.~A.~Lauber, with L.~L\"onnblad and M.~Seymour. These Proceedings.
%-----------------------------------------------------------------------
\bibitem{OPALZeits}
K.~Ackerstaff et~al.: Z. Phys. C74 (1997) 33--48, OPAL
  Collaboration
%-----------------------------------------------------------------------
\bibitem{Blobel}
V.~Blobel, DESY-84-118, 1984.
%-----------------------------------------------------------------------
\bibitem{Kartvili}
A.~H\"ocker and V.~Kartvelishvili.  Nucl. Instr. Meth. {\bf A372} (1996) 469
%-----------------------------------------------------------------------
\bibitem{TASSO}
TASSO collaboration, M.~Althoff {\em et al}, Z.Phys. {\bf C31} (1986) 527
%-----------------------------------------------------------------------
\bibitem{TPC}
$TPC/2\gamma$ collaboration, H.~Aihara {\em et al}, Z.Phys. {\bf C34} (1987) 1
%-----------------------------------------------------------------------
\bibitem{Vogt}
A.~Vogt.  These Proceedings
%-----------------------------------------------------------------------
\bibitem{GRV}
M.~Gl{\"u}ck, E.~Reya and A.~Vogt, Phys. Rev. {\bf D46} (1992) 1973\\
M.~Gl{\"u}ck, E.~Reya and A.~Vogt, Phys. Rev. {\bf D45} (1992) 3986
%-----------------------------------------------------------------------
\bibitem{SaS-1D}
G.~A.~Schuler and T.~Sj{\"o}strand, Z. Phys. {\bf C68} (1995) 607.
%-----------------------------------------------------------------------
\bibitem{ZEUSfirst} ZEUS Collaboration, M. Derrick {\it et
al.}, Phys. Lett. {\bf B316} (1993) 412.
%-----------------------------------------------------------------------
\bibitem{H1first} H1 Collaboration, I. Abt {\it et al.},
Nucl. Phys. {\bf B407} (1993) 515.
%-----------------------------------------------------------------------
\bibitem{Forshaw95}
J.~R.~Forshaw.  Proceedings of Photon '95, Sheffield, England 8-13 April 1995,
eds Miller, Cartwright and Khoze;  World Scientific, Singapore (1995) 3.
%-----------------------------------------------------------------------
\bibitem{HERWIG}
G.~Marchesini and B.~R.~Webber, Nucl. Phys. {\bf B310} (1988) 461
M.~H.~Seymour, Z. Phys {\bf C56} (1992) 161
%-----------------------------------------------------------------------
\bibitem{LauberWw}
J.~A.~Lauber for the OPAL collaboration.  Proceedings of ICHEP '96, Warsaw,
eds Z.~Ajduk and A.~K.~Wroblewski, World Scientific, Singapore (1996) 725
%-----------------------------------------------------------------------
\bibitem{PYTHIA}
T.~Sj{\"o}strand, Comp. Phys. Comm. {\bf 82} (1994) 74\\
T.~Sj{\"o}strand, PYTHIA 5.7 and JETSET 7.4: Physics and Manual,
CERN-TH/93-7112
%-----------------------------------------------------------------------
\bibitem{ARIADNE}
L.~L\"onnblad.  Comp. Phys. Comm. {\bf 71} (1992) 15.
%-----------------------------------------------------------------------
\bibitem{DJMLund}
D.~J.~Miller.  Proceedings of Workshop on Two-Photon Physics at LEP and HERA,
Lund, Sweden, May 1994; eds G.~Jarlskog and L.~J\"onsson, Lund University
(1994) 4
%-----------------------------------------------------------------------
\bibitem{F2GEN}
A.~Buijs, W.~G.~J.~Langeveld, M.~H.~Lehto and D.~J.~Miller.  Comp. Phys. Comm.
{\bf79} (1994) 523
%-----------------------------------------------------------------------
\bibitem{Rooke}
A.~M.~Rooke for the OPAL collaboration.  These Proceedings.
%-----------------------------------------------------------------------
\bibitem{Zimin}
N.~Zimin for the DELPHI collaboration.  These Proceedings.
%-----------------------------------------------------------------------
\bibitem{Engel}
R.~Engel, with F.~W.~Bopp, J.~Ranft and A.~Rostovtsev.  These Proceedings.
%-----------------------------------------------------------------------
\bibitem{Buergin}
R.~B{\"u}rgin for the OPAL collaboration.  These Proceedings.
%-----------------------------------------------------------------------
\bibitem{Witten}
E.~Witten, Nucl. Phys.{\bf B120} (1977) 189
%-----------------------------------------------------------------------
\bibitem{Millergreenbook}
D.~J.~Miller.  Proceedings of ECFA workshop on Physics at LEP200, Aachen 1986,
eds A.~B{\"o}hm and W.~Hoogland; CERN 87-08; ECFA 87/108; 207
%-----------------------------------------------------------------------
\bibitem{Field95}
J.~H.~Field.   Proceedings of Photon '95, Sheffield, England 8-13 April 1995,
eds Miller, Cartwright and Khoze;  World Scientific, Singapore (1995) 485.
%-----------------------------------------------------------------------
\bibitem{LEP2wkshp}
P.~Aurenche {\em et al}; $\gamma \gamma$ working group report; in Physics at
LEP2, CERN 96-01, 301
%-----------------------------------------------------------------------
\bibitem{Brew}
C.~A.~Brew for the ALEPH collaboration.  These Proceedings.
%-----------------------------------------------------------------------
\bibitem{Doucet}
M.~Doucet for the OPAL collaboration.  These Proceedings.
%-----------------------------------------------------------------------
\bibitem{Rick}
H.~Rick for the H1 collaboration.  These Proceedings.
%-----------------------------------------------------------------------
\bibitem{Erdmann}
M.~Erdmann for the H1 collaboration.  Proceedings of Photon '95, Sheffield,
England 8-13 April 1995, eds Miller, Cartwright and Khoze;  World Scientific,
Singapore (1995) 59.
%-----------------------------------------------------------------------
\bibitem{Butterworth}
J.~M.~Butterworth for the ZEUS collaboration.  Proceedings of Photon '95,
Sheffield, England 8-13 April 1995, eds Miller, Cartwright and Khoze;  World
Scientific, Singapore (1995) 53
%-----------------------------------------------------------------------
\bibitem{MillerCornell}
D.~J.~Miller.  Proceedings of the Lepton Photon Symposium, Cornell University,
August 1993, eds P.~Drell and D.~Rubin, AIP Conference Proceedings No. 302; 654
%-----------------------------------------------------------------------
\bibitem{ChrLls}
C.~ H.~Llewellyn-Smith, Phys. Lett. {\bf B79} (1978) 83.
%-----------------------------------------------------------------------
\bibitem{Aurenche}
P.~Aurenche.  These Proceedings.
%-----------------------------------------------------------------------
\bibitem{KandK}
T.~Kleinwort and G.~Kramer.  Phys. Letts {\bf B370} (1996) 141
%-----------------------------------------------------------------------
\bibitem{LAC}
H.~Abramowicz, K.~Charchula and A.~Levy, Phys. Lett. {\bf B269} (1991) 458
%-----------------------------------------------------------------------
\bibitem{Pancheri}
G. Pancheri. These Proceedings.
%-----------------------------------------------------------------------
\bibitem{VanRossum}
W.~van~Rossum for the L3 collaboration. These Proceedings.
%-----------------------------------------------------------------------
\bibitem{Andreev}
V.~P.~Andreev for the L3 collaboration. These Proceedings.
%-----------------------------------------------------------------------
\bibitem{DKZZ}
M.~Drees, M.~Kr\"amer, J.~Zunft and P.~M.~Zerwas, Phys. Lett. {\bf B306}
(1993) 371
%-----------------------------------------------------------------------
\bibitem{Hautmann}
F.~Hautmann with S.~Brodsky and D.~Soper.  These Proceedings
and Phys. Rev. Lett. {\bf 78} (1997) 803
%-----------------------------------------------------------------------
\bibitem{deRoeck}
J.~Bartels, A.~De~Roeck and H.~Lotter.  DESY-96-168, hep-ph/9608401
%-----------------------------------------------------------------------
\bibitem{Brodsky}
S.~Brodsky.  These Proceedings
%-----------------------------------------------------------------------
\bibitem{Savinov}
V.~Savinov for the Cleo II collaboration.  These Proceedings.
%-----------------------------------------------------------------------
\bibitem{Braccini}
S.~Braccini for the L3 collaboration.  These proceedings.
%-----------------------------------------------------------------------
\bibitem{Ruskov}
R.~Ruskov with A.~V.~Radyushkin.  These Proceedings.
%-----------------------------------------------------------------------
\bibitem{Paar}
H.~Paar for the Cleo II collaboration.  These Proceedings.
%-----------------------------------------------------------------------
\bibitem{Chanowitz}
M.~S.~Chanowitz.  proceedings of the VIIIth International Workshop on
Photon-Photon Collisions, Shoresh, Israel, April 1988; ed U.~Karshon.  World
Scientific, Singapore (1988) 205
%-----------------------------------------------------------------------
\bibitem{Medin}
G.~Medin for the ARGUS collaboration.  These Proceedings.
%-----------------------------------------------------------------------
\bibitem{Tapprogge}
S.~Tapprogge for the H1 collaboration.  These Proceedings.
%-----------------------------------------------------------------------
\bibitem{Romanov}
L.~Romanov for the KEDR collaboration.  These Proceedings.
%-----------------------------------------------------------------------
\bibitem{Maslennikov}
A.~Maslennikov for the KEDR collaboration.  These Proceedings.
%-----------------------------------------------------------------------
\bibitem{Telnov}
V.~Telnov.  These Proceedings.
%-----------------------------------------------------------------------
\bibitem{Jikia}
G.~Jikia.  These Proceedings
%-----------------------------------------------------------------------
\bibitem{Ginzburg}
I.~F.~Ginzburg.  These Proceedings
%-----------------------------------------------------------------------
\bibitem{Takahashi}
T.~Takahashi.  These Proceedings.
%-----------------------------------------------------------------------
\bibitem{DJMMorioka}
D.~J.~Miller.  Proceedings of the Workshop on Physics and Experiments with
Linear Colliders, Morioka-Appi, Japan, September 1995, eds A.~Miyamoto,
Y.~Fujii, T.~Matsui, S.~Iwata, World Scientific, Singapore (1996) 322
%-----------------------------------------------------------------------
\bibitem{VogtMiller}
D.~J.~Miller and A.~Vogt. $e^+e^-$ Collisions at TeV Energies; the Physics 
Potential.  ed P.M.Zerwas, DESY 96-123D (1996) 473
%-----------------------------------------------------------------------
\bibitem{Borden}
D.~A.~Bauer, D.~L.~Borden, D.~J.~Miller and J.~E.~Spencer, SLAC-PUB-5816 (1992)
%-----------------------------------------------------------------------
\bibitem{MillerBerk}
D.~J.~Miller. Nucl. Instr. Meth. {\bf A355} (1995) 101
\end{thebibliography}
\end{document}